\documentclass[aps,prl,preprint]{revtex4-1}

\usepackage{graphicx}  
\usepackage{amsmath} 
\usepackage{dcolumn}   
\usepackage{bm}        
\usepackage{amssymb}   
\usepackage{epstopdf}
\usepackage{xcolor} 
\usepackage[normalem]{ulem}

\def \Vtg {V$_{TG}$}
\def \Vbg {V$_{BG}$}
\def \Rxx {R$_{xx}$~}
\def \Rxy {R$_{xy}$~}
\def \BSTS {BiSbTe$_{1.25}$Se$_{1.75}$}
\def \Rq {$h/e^2$}
\def \Rqq {h/e^2}
\begin{document}

\title{Quantized transport in topological insulator n-p-n junctions} 

\author{Abhishek Banerjee$^1$, Ananthesh Sundaresh$^1$, Sangram Biswas$^1$, R. Ganesan$^1$, 
Diptiman Sen$^2$, and P. S. Anil Kumar$^1$}
\affiliation{\small{$^1$Department of Physics, Indian Institute of Science, 
Bengaluru 560012, India \\
$^2$Centre for High Energy Physics, Indian Institute of Science, Bengaluru
560012, India}}


\begin{abstract}
Electrical transport in three dimensional topological insulators(TIs) occurs through spin-momentum locked topological surface states that enclose an insulating bulk. In the presence of a magnetic field, surface states get quantized into Landau levels giving rise to chiral edge states that are naturally spin-polarized due to spin momentum locking. It has been proposed that p-n junctions of TIs in the quantum Hall regime can manifest unique spin dependent effects, apart from forming basic building blocks for highly functional spintronic devices. Here, for the first time we study electrostatically defined n-p-n junctions of bulk insulating topological insulator \BSTS~in the quantum Hall regime. We reveal the remarkable quantization of longitudinal resistance into plateaus at 3/2 and 2/3 $\Rqq$, apart from several partially developed fractional plateaus. Theoretical modeling combining the electrostatics of the dual gated TI n-p-n junction with Landauer Buttiker formalism for transport through a network of chiral edge states explains our experimental data, while revealing remarkable differences from p-n junctions of graphene and two-dimensional electron gas systems. Our work not only opens up a route towards exotic spintronic devices but also provides a test bed for investigating the unique signatures of quantum Hall effects in topological insulators. 
\end{abstract}

\maketitle
\subsection{Introduction}
Three dimensional topological insulators have drawn intense interest as materials that host gapped insulating bulk states but topologically protected conducting states on the surface~\cite{TIreview1,TIreview2,TIreview3,TIreview4,TIreview5,TIreview6}. The topological surface states(TSS) consist of spin-helical Dirac fermions with the direction of electronic spin locked to the direction of momentum. Recent advances in material technology have allowed TSS to be robustly accessed and manipulated, spawning a gamut of experimental breakthroughs including the quantum anomalous Hall effect~\cite{QAHE1}, topological magnetoelectric effect~\cite{axion}, observation of Majorana fermions~\cite{Majorana_TI} and the half-integer quantum Hall effect~\cite{QHTI2, Half-integer-QHE2}. Specifically, the quantum Hall effect in topological insulators arising from a combination of the top and bottom topological surface states has been recently measured~\cite{QHTI1, QHTI2, QHTI3, QHTI4}. Transport results have been interpreted as arising from the intriguing half-integer quantization of Dirac fermions, however direct signatures of the {\it topological} nature of transport of chiral edge states in these systems is lacking. The quantum Hall effect in 3D TIs is quite different from conventional two-dimensional electron gas(2DEG) systems and even graphene. Unlike the latter systems, surface states in 3D TIs have no real boundary and essentially enclose a 3D bulk~\cite{QHE-TI-1, QHE-TI-2, QHE-TI-3}. The conventional edge state picture of quantum Hall effect therefore does not apply for TIs; instead chiral edge modes appear at boundaries separating regions with different Landau level filling factors. More intriguingly, the lack of spin-(or valley) degeneracy and the unique spin-momentum locked structure of surface states results in spin-polarization of the low energy quantum Hall edge states~\cite{ilan2015spin}, resulting in physics that is unique to TIs. 

\begin{figure}[!t]
\includegraphics[width=1.\linewidth]{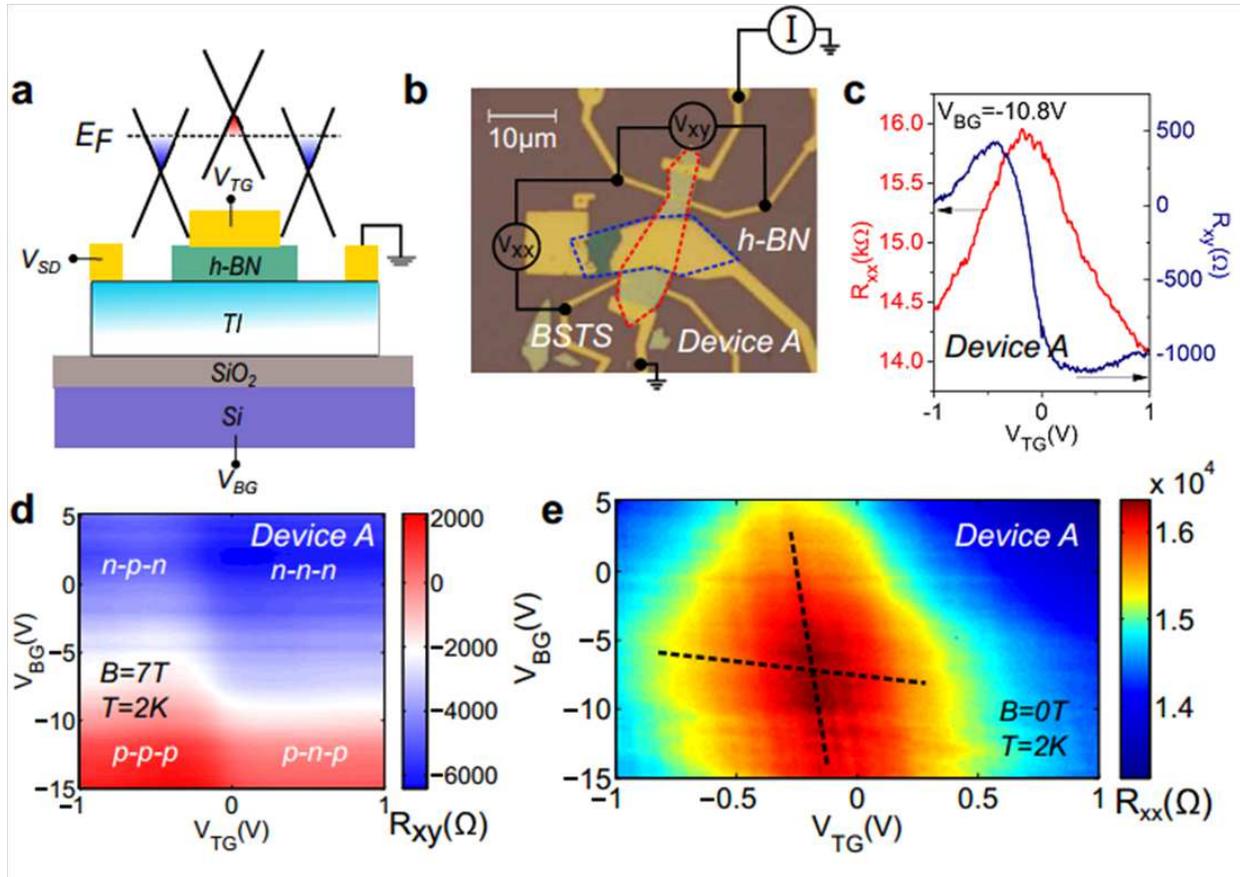}
\caption{(a) Schematic of a topological insulator n-p-n structure formed by a combination of local top gating and global back gating. (b) Optical micrograph of a typical device. Red and blue dotted lines represent the outlines of the BSTS and h-BN flakes respectively. (c) R$_{xx}$ (left) and R$_{xy}$ (right) as a function of \Vtg~at a fixed \Vbg=-10.8V. R$_{xx}$ measured at zero magnetic field, R$_{xy}$ measured at B=7T. (d) 2D color map of Hall resistance R$_{xy}$ (B=7T) and (e) \Rxx (B=0T) as a function of \Vtg~ and \Vbg~. The `vertical' and `horizontal' dotted lines indicate the charge neutrality points of the bottom and top topological surface states respectively.}
\label{fig01}
\end{figure} 

Previous works have shown that unlike samples with uniform carrier densities, p-n junction are highly sensitive to the spin-momentum locked structure of the surface state dispersion~\cite{PhysRevLett.114.176801, ilan2015spin, TIpn1, TIpn2}. Because n and p type regions on a TI surface posses opposite spin-momentum helicities, a spin filtering effect at the p-n interface gives rise to enhanced resistances~\cite{TIpn2}. This effect becomes even more dramatic in the quantum Hall regime, where spin-polarization of edge states emanating from the zero Landau level become fully locked to the direction of edge state propagation, giving rise to large spin-mixing resistances when edge states with different spin polarizations mix at the p-n interface, constituting a perfect spin-filter~\cite{ilan2015spin}. In fact, in the presence of in-plane magnetic fields, such devices comprise an ideal realization of the Datta-Das spin-FET~\cite{ilan2015spin,datta-das}. TI p-n junctions in the quantum Hall regime can therefore be utilized to study various unique aspects of quantum Hall effect in TIs, apart from providing a lucrative platform for building gate-tunable spintronic devices. In this work, for the first time we study n-p-n junctions of the bulk insulating topological insulator \BSTS. We employ a dual-gated geometry where the top gate partially covers the top surface of the sample, while a global back gate controls the carrier density in the rest of the device forming a double-junction structure. In the presence of a large perpendicular magnetic field, chiral edge states appear at the boundaries separating regions of the sample with different Landau level indices. Transport across a network of one dimensional chiral edge states gives rise to the unique observation of fractional resistance plateaus at 3/2 and 2/3 $\Rqq$ close to the zero Landau level. We also observe several other partially formed fractional plateaus. We construct a model based on the Landauer Buttiker formalism for edge state propagation that quantitatively explains our data. We show that the nature of quantum Hall edge state formation and mixing in TI p-n junctions is remarkably different from that in graphene and 2DEG p-n junctions, leading not only to qualitative, but also quantitative distinctions.

\subsection{Results}
\subsubsection{\bf Dual gated field-effect transport}
Our devices are fabricated from exfoliated thin flakes of the topological insulator \\\BSTS~(BSTS), belonging to the class of highly bulk insulating TIs~\cite{insulatingTI1, insulatingTI2, insulatingTI3, BSTS2}. Previous angle-resolved photoemission spectroscopy (ARPES) studies~\cite{lohani2017} and transport experiments~\cite{banerjee2016} by us show that the chemical potential in this material lies within the bulk band gap, and the bulk conduction is undetectable below 100K. A thin layer of hexagonal Boron-Nitride (h-BN) covering the BSTS flake serves as the top gate dielectric while the Si/SiO$_2$ substrate forms the back gate (Fig.~\ref{fig01}(a) and (b)). The h-BN flake is metallized in the central region to obtain a locally top gated region. A combination of local and global gates allows us to reconfigurably tune the position of the chemical potential to different values in adjacent regions of the sample as shown Fig.~\ref{fig01}(a). Several such devices have been fabricated and measured. Here we present data from two of them, device A and B. Measurements are performed in a 2K cryostat with a 7T magnet, and 25mK dilution fridge with 16T magnet. Resistance measurements are carried out in current biased 4-terminal configuration(Fig.~\ref{fig01}(b)) using standard low-frequency lock-in techniques.

The double gated field-effect behavior measured at T=2K is shown in Fig.~\ref{fig01}(c) for device A. For a given fixed back-gate voltage
(\Vbg = -10.8V), the variation of the top-gate voltage in a small window of -1V to 1V is found to be sufficient to drive the chemical potential across the charge neutrality point (CNP) of the top surface state. This is represented as a maximum in the longitudinal resistance R$_{xx}$, corresponding to the minimum density of states at the Dirac point. The corresponding measurement of the Hall resistance, R$_{xy}$, at a magnetic field of 7T shows that the carrier type switches from p-type to n-type. Similarly, the bottom surface state chemical potential can be tuned using the bottom gate electrode. A combination of top and bottom gates can therefore be used to access ambipolar transport in our devices. This is exemplified in the color plot of Fig.~\ref{fig01}(d) where the variation of R$_{xy}$ as a function of both \Vtg~ and \Vbg~ shows clear regions of n-type (blue) and p-type (red) conduction. The corresponding plot for \Rxx at zero magnetic field (Fig.~\ref{fig01}(e)) shows a characteristic rhombus shape, where the horizontal and vertical diagonals represent the CNPs of the top and bottom surface states respectively. The exact trajectories of the CNPs can be determined from universal conductance fluctuation (UCF) measurements described in the supplementary material section D. The charge neutrality peak at the intersection of the diagonals at (\Vtg,\Vbg)=(-0.12V,-6.2V) corresponds to the overall CNP of the sample. From this we estimate the initial electron densities on the top and bottom surfaces to be $n_T^0=2.6\times10^{11}/cm^2$ and $n_B^0=3\times10^{11}/cm^2$. The tilting of the diagonals away from the horizontal and vertical directions indicates a finite capacitive coupling between the top gate electrode and the bottom surface and vice versa. This coupling is mediated by the highly insulating bulk of the BSTS flake, acting like a capacitor, and it plays a central role in our experiments. Additional characterization of our devices including measurements of surface state coupling parameters from weak anti-localization analysis, temperature and gate-bias dependent Hall effect measurements and UCF analysis are provided in the supplementary material sections B, C and D.
 
\subsubsection{\bf Quantization of longitudinal resistance}  
\begin{figure}[!t]
\includegraphics[width=1\linewidth]{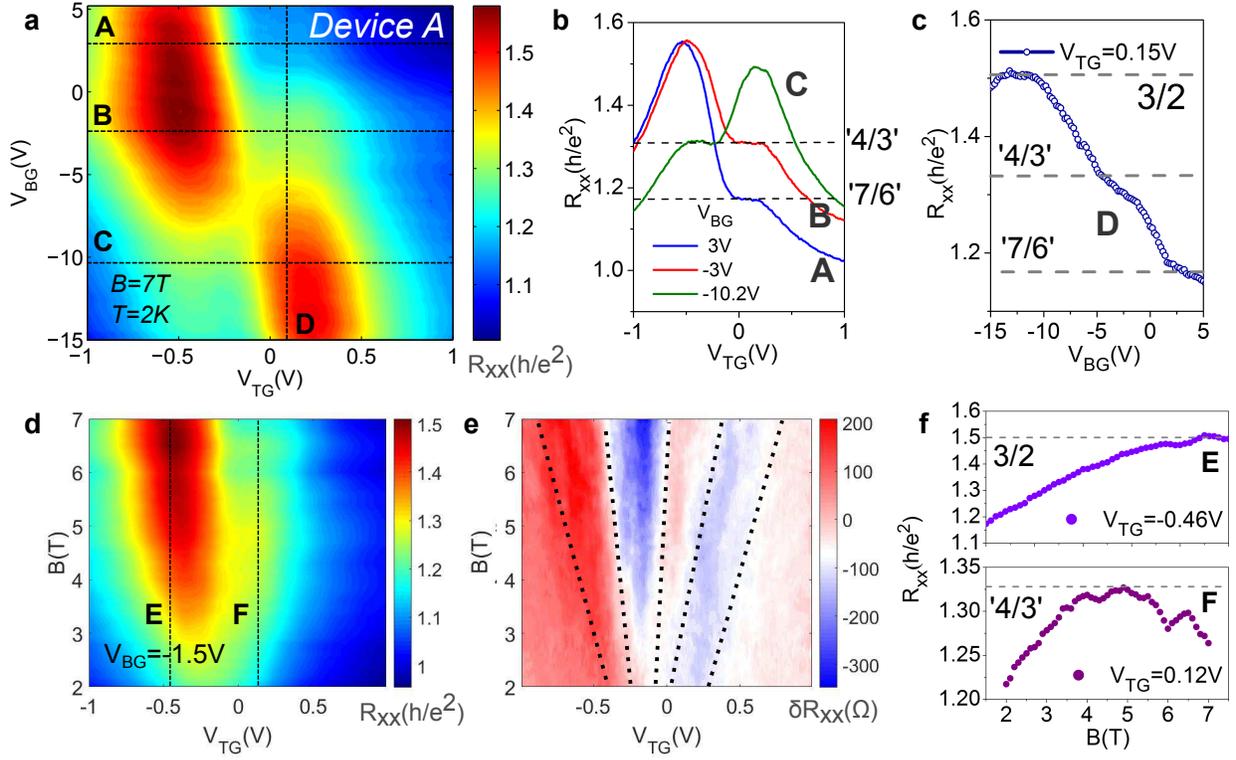}
\caption{(a) 2D color map of \Rxx as a function of \Vbg~ and \Vtg~ at B=7T. (b) \Rxx as a function of top gate voltage at \Vbg=3V, -3V and -10.2V. Quantized fractional plateaus appear at \Rxx=4/3 and 7/6 \Rq. (c) \Rxx as a function of back-gate voltage at fixed top gate voltage \Vtg=0.15V. A distinct plateau appears at \Rxx=3/2 \Rq, whereas the two other plateaus at \Rxx=4/3 and 7/6 are less well-developed.(d) \Rxx vs \Vtg~as a function of magnetic field measured at \Vbg=-1.5V (e) Landau fan plot obtained by plotting the first difference d\Rxx vs \Vtg~as a function of magnetic field (f) Trajectories of the `3/2'(upper panel) and `4/3'(lower panel) plateaus as a function of magnetic field.}
\label{fig02}
\end{figure}

We now turn to the longitudinal resistance measurements in quantizing magnetic fields, which forms the bulk of this work. The color plot in Fig.~\ref{fig02}(a) shows \Rxx measured at temperature of T=2K and magnetic field B=7T applied perpendicular to the top and bottom surfaces of device A. The resistance map exhibits a rich pattern consisting of different regions where \Rxx shows quantized values. The pattern is symmetric about the overall CNP at (\Vtg,\Vbg)=(-0.12V,-6.2V) extracted in Fig.~\ref{fig01}(e). The most distinctive feature is the splitting of the single charge neutrality peak at zero magnetic field (Fig.~\ref{fig01}(e)) to two large symmetric plateaus with resistance $\sim$ 3/2 \Rq. This large plateau is flanked by two smaller features at (4/3) \Rq and (7/6) \Rq as shown in Fig.~\ref{fig02}(b), where the \Vtg~ linescans at \Vbg=3V, -3V and -10.2V correspond to the horizontal slices labeled A, B and C respectively in Fig.~\ref{fig02}(a). As seen in Fig.~\ref{fig02}(b), the remarkable flatness of \Rxx at the (4/3) \Rq and (7/6) \Rq (slices A and B) are indicative of quantum Hall behavior. At the same time, a comparison of slices B and C which are measured at roughly $\Delta$\Vbg$ \simeq \pm(3-4V)$ away from the CNP in either direction shows that 4/3 \Rq~plateau is reflected symmetrically across \Vtg$=0V$, indicating a well-maintained particle-hole symmetry. The peaks in \Rxx at 1.5-1.6 \Rq do not appear to be well-quantized in the \Vtg~ linescans. On the other hand, the \Vbg~ linescan data clearly demonstrates the quantization of the 3/2 plateau as seen in Fig.~\ref{fig02}(c). In turn, the features at (4/3) \Rq and (7/6) \Rq appear less quantized. 

 \begin{figure}[!t]
\includegraphics[width=1\linewidth]{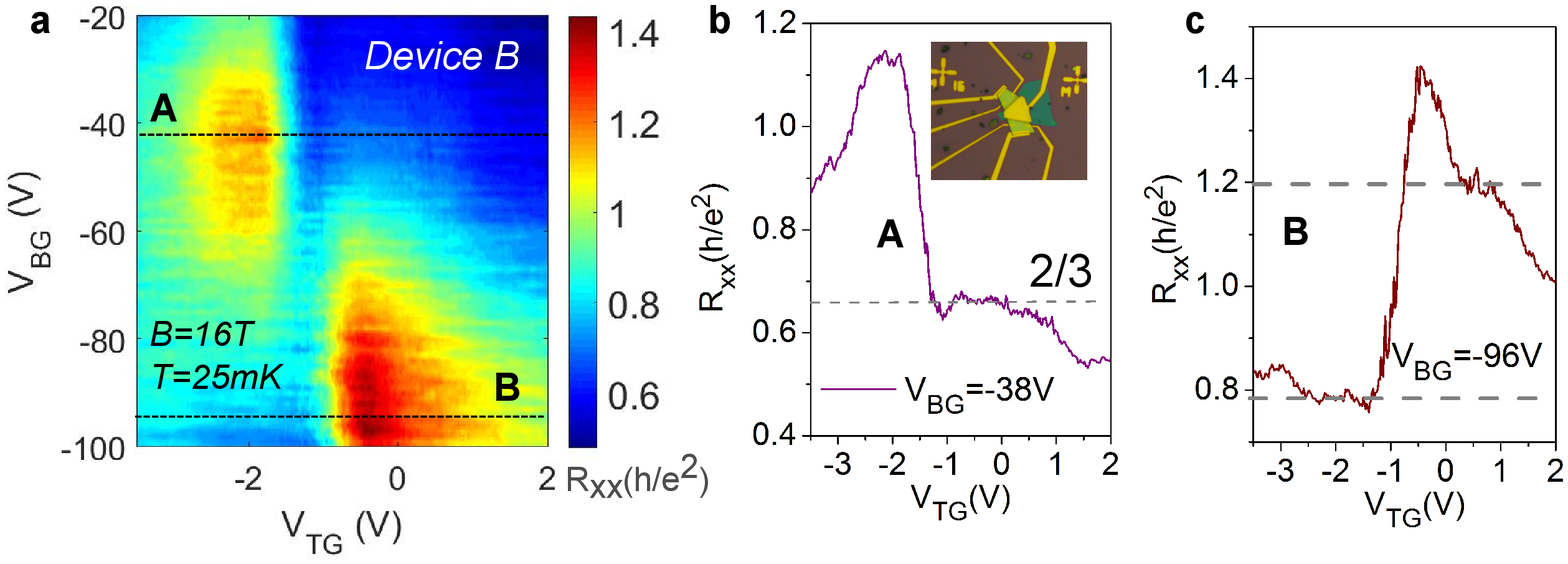}
\caption{(a) 2D color map of \Rxx as a function of \Vbg~ and \Vtg~ at B=16T. (b) \Rxx as a function of top gate voltage at \Vbg=-38V and (c) \Vbg=-96V. Quantized fractional plateau appears at \Rxx=2/3 \Rq. Sample temperature is 25mK}
\label{fig03}
\end{figure} 

To verify whether these plateau like features are fully or only partially quantized, we study the evolution of \Rxx at \Vbg=-1.5V as a function of increasing magnetic field as shown in Fig.~\ref{fig02}(d). In Fig.~\ref{fig02}(e), we plot the first difference of this data, showing a clear Landau fan diagram structure. This proves that data presented in Fig.~\ref{fig02}(a) is indeed a consequence of quantum Hall effect in our samples. In Fig.~\ref{fig02}(f), we plot the trajectories of the `3/2' and `4/3' features as a function of magnetic field. While the `3/2' plateau indeed saturates to 3/2 \Rq~as a function of magnetic field showing complete quantization, the feature at `4/3' clearly deviates from its `quantized' value. This is highly reminiscent of similar features observed recently in Cobalt doped BiSbTeSe$_2$ samples~\cite{Half-integer-QHE2}, where a quantization of the Hall conductance at an anomalous half integer value $\sigma_{xy}=3/2 e^2/h$ was observed only at intermediate fields($\sim$ 7T) and subsequently vanished at higher magnetic fields. This was explained on the basis of a delayed Landau level hybridization model involving half-integer quantum Hall effect of a single Dirac fermion surface state~\cite{Half-integer-QHE1}. We discuss a possible connection of the `4/3' feature with the half-integer quantum Hall effect later. 

Similar experiments are performed on device B at a larger magnetic field $B=16T$ and lower temperatures $T=25mK$. Fig.~\ref{fig03}(a) shows a plot of the longitudinal resistance measured as a function of \Vtg~ and \Vbg. Similar to device A, device B also shows a splitting of the peak at the charge neutrality point(\Vbg $\sim$ -60V, \Vtg $\sim$ -1.5V) into two peaks along the diagonal with \Rxx $\sim$ 3/2 \Rq~ for the lower peak at \Vbg $<$-60V. More strikingly, we observe a strongly quantized plateau at \Vbg $\sim$ -38V where \Rxx=2/3 \Rq(Fig.~\ref{fig03}(b). The simultaneous appearance of a plateau in the Hall resistance data at \Rxy=1/3 \Rq~confirms that the 2/3 plateau is fully quantized(see supplementary material). Note that in device A, we observe a partially developed plateau at \Rxx = 4/3 \Rq~ at a similar region of the \Vtg-\Vbg~space in intermediate magnetic fields.

\subsubsection{\bf Theoretical model}

To understand these unusual fractional resistance plateaus, we develop a model of one-dimensional edge-state propagation in the quantum Hall regime for a topological insulator with strong carrier density gradients. Similar models have been used to analyze conductance quantization in high-mobility two-dimensional electron gas (2DEG) systems~\cite{haug1988} and recently in graphene~\cite{Graphene_npn1,Graphene_npn2,Graphene_npn3,Graphene_npn4}. While 2DEG systems allow only unipolar type n-n' or p-p' junctions, zero-gap semiconductors like graphene allow the formation of bipolar type n-p' and p-n' type junctions. Specifically, for the case of graphene, the quantum Hall effect in n-p-n junctions has been used to study a rich variety of physics, including edge-state equilibration~\cite{Graphene_npn1}, spin and valley polarization of quantum Hall edge states~\cite{Graphene_npn2}, and formation of snake states~\cite{snake1, snake2}, apart from providing a technologically important platform for applications. However, topological insulators differ from the previously considered systems in certain striking ways: i) Unlike a compact two-dimensional surface such as graphene or 2DEG, the surface states in a topological insulator lack a true boundary and instead completely enclose the 3D insulating bulk. In a slab geometry such as ours, the top and bottom topological surface states get quantized into Landau levels in the presence of a perpendicular magnetic field while the lateral surfaces that are parallel to the magnetic field host the one-dimensional quantum Hall edge states. The connection between the half-integer quantization of the top and bottom surface Landau levels, and the number of chiral edge states on the lateral surface is a matter of active investigation~\cite{QHE-TI-1, QHE-TI-2, QHE-TI-3}. ii) The spin-momentum locking in topological insulators gives rise to naturally spin-polarized edge states, unlike graphene and other 2DEGs. While this effect is strongest for edge states arising from the lowest energy LLs, it can lead to dramatic changes to transport in p-n junction structures.~\cite{ilan2015spin} iii) The three dimensional nature of surface states makes it imperative to consider the detailed electrostatics of the system to account for the relatively large capacitive coupling between the top and bottom surface states due to the insulating bulk of the TI~\cite{gating3}. Therefore, the Landau level indices of the top and bottom TSS have to be calculated self-consistently.

\begin{figure}[!t]
\includegraphics[width=1.\linewidth]{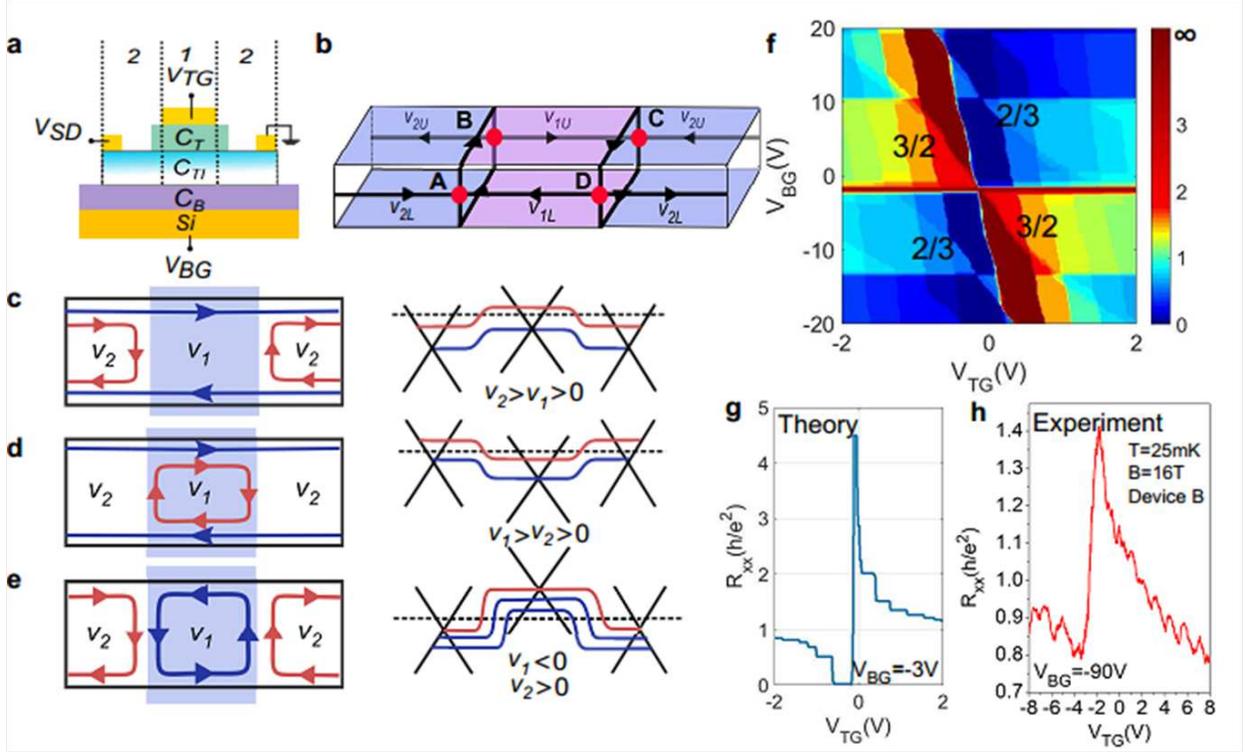}
\caption{(a) Schematic of the device structure showing region-1 (top gated) and region-2 (back-gated). (b) Edge state formation on the top and bottom topological surface state in the n-p-n regime. (c), (d) and (e) (Left) Two-dimensional view of edge state formation and equilibration in the unipolar and bipolar regimes. (Right) Corresponding Landau level diagrams.(f) Simulated 2D map of resistance quantization with sample dimensions $t_{hBN}=15nm$, $t_{SiO_2}=500nm$ and $t_{BSTS}=20nm$, representing the thicknesses of hBN, SiO$_2$ and the BSTS layers respectively. Magnetic field B=7T. The first and third quadrant correspond to the unipolar regime, second and fourth quadrants are the bipolar regime (g) and (h) Theoretical and experimental values of \Rxx~vs \Vtg~with \Vbg~set slightly lower than the charge neutrality point showing the crossover from unipolar to bipolar regime with increasing top gate voltage.}
\label{fig04}
\end{figure}

The dissimilar electrostatic boundary conditions on the top and bottom surface states (Fig.~\ref{fig04}(a)) breaks the structural inversion symmetry. Each surface consists of two distinct portions: the dual gated portion, region-1, and the back gated portion, region-2 which appears symmetrically on either side of the region-1 (Fig.~\ref{fig04}(a)). Considering both surfaces together, there are four distinct regions with unique filling factors $\nu_{j\sigma}$ with $j=1, 2$ and $\sigma=U,L$; $j$ representing the region number and $\sigma$ is the layer index. (Fig.~\ref{fig04}(b)). The geometrical capacitances of the top gate($C_T$), bottom gate($C_B$) and TI bulk ($C_{TI}$) and the two gate voltages completely determine the filling factors on each surface of the sample. The connection between Landau level filling factors on top and bottom surfaces and the number of chiral edge states that propagate along the lateral surface has been evaluated in previous works~\cite{QHE-TI-1, QHE-TI-2, QHE-TI-3} . In the presence of structural inversion asymmetry(SIA) generated by the potential difference between the top and bottom surfaces, an integer number of chiral edge modes given by $\nu_{j}=\nu_{j,U}+\nu_{j,L}$, where j=1,2 appear along the sidewall in each region as shown in Fig.~\ref{fig04}(b)~\cite{QHTI2, QHE-TI-2}. A separate set of edge modes that traverse the two sample surfaces in the transverse direction appear at the two p-n interfaces. These edge modes appear as Landau levels in regions with different carrier densities cut across the chemical potential(right panel of Figs.~\ref{fig04}(c)-(e)). The number of chiral edge states flowing at the p-n interfaces on each surface(U and L) is given as $\nu_\sigma=\nu_{2,\sigma}-\nu_{1,\sigma}$  where $\sigma=U,L$. The total number of transverse edge modes separating region-1 and region-2 are therefore given as $\nu_U+\nu_L$=$\nu_1 -\nu_2$. This reduces the three-dimensional network of edge states shown in Fig.~\ref{fig04}(b) to an effective two-dimensional model characterized completely by $\nu_1$ and $\nu_2$(Figs.~\ref{fig04}(c)-(e)).

Previous work~\cite{ilan2015spin} has shown that each of these edge states are spin-polarized and therefore spin-mixing resistances need to be taken into account at the vertices where four edge modes meet(A, B, C and D in Fig.~\ref{fig04}(b)). While this can lead to fascinating interferometric effects~\cite{ilan2015spin}, we neglect the spin mixing effects in this discussion. Details of derivation are provided in the supplementary material section F and G. In the quantum Hall regime, the conductance depends on the Landau level filling factors under and outside the top-gated region, $\nu_1$ and $\nu_2$ respectively. In Figs~\ref{fig04}(c)-(e), we present a two-dimensional projection of the edge state network appearing in three distinct regimes. In the unipolar regime, $\nu_1$ and $\nu_2$ have the same sign. Here, two distinct cases are possible. When $\nu_1 < \nu_2$, only $\nu_1$ number of edge states that are common to both the regions are allowed to transmit. The excess modes outside the gated region reflect back as shown in Fig.~\ref{fig04}(c) (Left) and the four terminal resistance is given by: $R=(\Rqq) \left(1/|\nu_1|-1/|\nu_2|\right)$. When $\nu_1 > \nu_2$, the gated region contains excess quantum Hall states that circulate around its edges as shown in Fig.~\ref{fig04}(d) (Left). Along the edges of the sample, the edge states from the un-gated region and the excess edge states from the gated region mix together and equilibrate, giving $R=(\Rqq) \left(1/|\nu_2|-1/|\nu_1|\right)$. In both these cases, we find that the four-terminal longitudinal resistance gets quantized at fractional values:
\begin{equation}
\frac{R}{(\Rqq)} =\left|\frac{1}{|\nu_2|}-\frac{1}{|\nu_1|}\right|=\frac{1}{2}, \frac{2}{3}, \frac{3}{4}, \cdots,
\end{equation}
where $\nu_1\nu_2>0$ and $\nu_1, \nu_2=1,2,3,\dots$.

The most interesting scenario is the bipolar regime depicted in Fig.~\ref{fig04}(e) where the edge states in the top-gated and un-gated regions circulate in opposite directions, and are localized to the respective region boundaries. The only way an electron can travel from left to right is by traversing along the transverse edge states flowing along the top/bottom surfaces of the sample, giving:
\begin{equation}
\frac{R}{(\Rqq)} = \left(\frac{1}{|\nu_2|}+\frac{1}{|\nu_1|}\right)=2, \frac{3}{2}, \frac{4}{3}, \cdots,
\end{equation}
where $\nu_1\nu_2<0$ and $\nu_1, \nu_2=1,2,3,\dots$. Note that these values of fractional resistances are distinctly different from those obtained in graphene~\cite{Graphene_npn1,Graphene_npn2} where $\nu=2, 6, 10,\dots$.

\subsubsection{\bf Comparison with experiment}

By solving the electrostatics of the dual gated TI n-p-n junction and combining it with the formula for the quantized resistance (Eqs. 1 and 2, see Supplementary material), we can now simulate resistance maps on each of the two surfaces of the sample. Fig.~\ref{fig04}(f) shows a 2D map of the top surface state resistance. Comparison with the experimental data of Fig.~\ref{fig02}(a) and Fig.~\ref{fig03}(a) shows a striking match between the positions and shapes of various plateaus. A distinct 3/2 plateau is observed in the bipolar regime, exactly as in our experiments, generated by $\nu_1=1$ and $\nu_2$=1/2, corresponding to the two lowest filling factors. The 2/3 plateau observed in the unipolar regime (Fig.~\ref{fig03}(b)) is also reproduced in our theoretical map as a dominant plateau. Also, as evident from Fig.~\ref{fig04}(f), parallel to the \Vtg-axis, the quantized regions appear as a quick succession of narrowly spaced plateaus, but parallel to the \Vbg-axis, they appear as large distinct plateaus. This is exactly as observed in our experiments (Fig.~\ref{fig02}(a),(c), Fig.~\ref{fig03}(a)). In fact, the quantization sequence obtained theoretically shown in Fig.~\ref{fig04}(g) and that obtained experimentally in Fig.~\ref{fig04}(h) (for \Vbg$<0$) show a striking qualitative match. 

However, we also note that several plateaus do not appear in our experiments, although they do appear in the model. Among these, the prominent plateaus arising from low energy Landau levels are \Rxx=2 \Rq~ in the bipolar regime and \Rxx=1/2 \Rq~ in the unipolar regime. Also we note that the extremal plateaus \Rxx$=0$($\nu_1=\nu_2$) and \Rxx$=\infty$($\nu_1, \nu_2=0$) do not appear in our experiment. These plateaus may be absent for a variety of reasons including partial quantization of the Landau levels, residual conduction through three dimensional bulk states, inter-Landau level hopping and the presence of residual Rashba-type topologically trivial surface states. In fact, our previous work indicates sizable conduction through Rashba states that live on the sample surface~\cite{banerjee2016} and may allow parasitic conduction between opposite edge states. However, another possibility is the lack of appearance of certain filling factors altogether. Although we assume the appearance of all integer filling factors in our model($\nu=0, 1, 2...$) because of structural inversion asymmetry, experiments on TI samples show varied behavior including only odd-integer filling fractions($\nu=1, 3,...$)~\cite{QHTI1}. This is attributed to `accidental' degeneracy between the top and bottom surface state Landau levels~\cite{QHE-TI-2}. Further, the character of the $\nu=0$ state is also hotly debated since it may either signify a complete absence of edge states or presence of dissipative one-dimensional edge states. Another tantalizing possibility explaining the discrepancies may have to do with the fact that we neglect altogether the spin mixing resistances in our calculations. This can lead to prefactors multiplying Eqs.1 and 2 and may lead to a resistance enhancement, especially for lower Landau levels. The prefactors are however difficult to evaluate and depend strongly on disorder, temperature and sample dimensions as shown in a previous work~\cite{ilan2015spin}.

\subsection{Discussion}
Our results on quantum transport in topological insulator n-p-n junctions are quite different from those obtained in 2DEG systems and graphene~\cite{Graphene_npn1, Graphene_npn2, Graphene_npn3, Graphene_npn4}. While the distinction from 2DEG p-n junctions is quite clear, in that, the reversal of transport characteristics(Figs.~\ref{fig02}(a) and ~\ref{fig03}(a)) across the charge neutrality point is not possible for a gapped semiconductor system, the differences with graphene are more subtle. In graphene samples, the number of edge states usually follows $\nu=4(n+1/2)=2,6,10,\cdots$ quantization~\cite{gr1, gr2, gr3, gr4}, the factor of 4 accounting for the valley and spin degeneracies giving rise to quite different fractional resistance sequences~\cite{Graphene_npn4, Graphene_npn1, Graphene_npn3}. Notably, the fractional resistances are constrained to be \Rxx $\leq$ 1 \Rq~in graphene, while we observe a clear 3/2 \Rq~plateau in our experiments. Only in ultra-clean samples and large magnetic fields are the valley and spin degeneracies broken, giving rise to the $\nu=1, 2, 3, \cdots$ integer quantization of Hall conductance~\cite{Graphene_npn2} where fractions larger than 1 \Rq~become possible. However in our experiments, the integer quantization can be achieved in comparatively `dirty' samples and low magnetic fields where Zeeman effects can be neglected. This opens up the possibility of utilizing the natural spin-polarization of TI edge states due to spin-momentum locking, which is otherwise destroyed by Zeeman effects. 

The low magnetic field regime is also interesting from the half-integer quantum Hall effect point of view~\cite{Half-integer-QHE1, Half-integer-QHE2,Half-integer-QHE3}. Although the quantization of a single Dirac fermion surface takes the half integer form($\nu=(n+1/2)$), the top and bottom surface states always appear as a pair and wash out the half-integer effect. However, recent experiments~\cite{Half-integer-QHE2} lend credence to the possibility that delayed quantization of one of the surface states with respect to the other(due to doping) at intermediate magnetic fields may lead to a non-trivial manifestation of the half-integer conductance quantization. The physical connection between $\sigma_{xy}=1/2$ and the number of edge states is not quite clear~\cite{Half-integer-QHE1, Half-integer-QHE2}, with theoretical works claiming that a measurement of the 1/2 integer QHE is not possible in a transport experiment~\cite{Half-integer-QHE3}. Nonetheless, we indicate that the `4/3' quantization observed in device A can be heuristically explained as a combination of the lowest Landau level `half-integer' edge states propagating only along the top surface state, $R=(1/\nu_1-1/\nu_2)=4/3$, with half-integer Hall conductances for gated and ungated regions, $\nu_1=1/2$ and $\nu_2=3/2$ respectively. As the magnetic field is increased, both surface states get completely quantized and the half-integer effect from a single surface disappears, producing a `2/3' quantization instead. 

Finally, we comment on the possibility of observing the unique spin-filtering effects produced by the spin-momentum locking of the chiral edge states. Such an effect would generically give rise to higher resistances compared with the `spinless' case we consider in our model. While certain disparities between theoretical and experimental resistances indeed indicate such an enhanced resistance effect, we cannot at present identify these effects uniquely with the spin-filtering effect. Samples with smaller dimensions and larger phase coherence lengths will be needed to observe these effects in future experiments. 

In conclusion, our work constitutes the first demonstration of quantum Hall effect in topological insulator n-p-n junctions. We open up a new avenue to study quantum Hall effect in Dirac systems without the restrictions imposed by degeneracy (spin and valley). Our work not only lays the foundation for studying the physics of spin-polarized chiral edge transport in TIs but also also paves the way towards high functional spintronic devices.

{\bf Methods}

{\bf Sample preparation}
Single crystals of \BSTS were grown by the modified Bridgman method using stoichiometric amounts of high purity(99.999\%) starting materials~\cite{banerjee2016}(see supplementary material section A). BSTS flakes are exfoliated using the scotch tape method onto heavily doped p-type silicon susbtrates with 285/500nm thick SiO$_2$ coating. Hexagonal Boron Nitride(h-BN) flakes of thickness $\sim$15-40nm are exfoliated from commercial h-BN powder(Momentive) and transfered onto the central portion of the BSTS flake using a dry Van der Waals transfer method, described in detail in section A of the supplementary material. Electrical contacts and the top-gate electrode are defined using e-beam lithography followed by e-beam evaporation of Cr/Au: 10nm/70nm.

{\bf Transport measurements}
Electrical transport measurements are performed using a low frequency AC lock-in technique with excitation currents of 10nA to 30nA in a helium-4 variable temperature cryostat(with base temperature of 1.7K, 8T magnetic field) or in a dry dilution refrigerator(with base temperature of 25mK, 16T magnetic field).

{\bf Acknowledgement}
A.B. thanks MHRD, Govt. of India for support. A.S. thanks KVPY, Govt. of India for support. D.S. thanks DST, India for support under Grant No. SR/S2/JCB-44/2010. P.S.A.K. thanks Nanomission, DST, Govt. of India for support.

\bibliography{Quantum_Hall_TI_bib_1}

\end{document}